\documentclass[letterpaper, 10 pt, conference]{IEEEtran}
\newcounter{mytempeqncnt}

\newcommand{\qed}{\nobreak \ifvmode \relax \else
      \ifdim\lastskip<1.5em \hskip-\lastskip
      \hskip1.5em plus0em minus0.5em \fi \nobreak
      \vrule height0.75em width0.5em depth0.25em\fi}

\ifCLASSINFOpdf
  \usepackage[pdftex]{graphicx}
  \graphicspath{{./}{Figs/}}
  \DeclareGraphicsExtensions{.pdf,.jpeg,.png}
\else
\fi
\usepackage{amssymb}
\usepackage{amsmath}
\usepackage{mathtools}
\usepackage{bigdelim}

\newcommand{\MeijerG}[7]{G^{#1,#2}_{#3,#4} \left(\! \begin{smallmatrix} #5 \\ #6 \end{smallmatrix} \middle\vert #7 \!\right) }

\IEEEoverridecommandlockouts

\begin{document}

\title{Opportunistic Spectrum Sharing using Dumb Basis Patterns: The Line-of-Sight Interference Scenario}
\author{\IEEEauthorblockN{Ahmed M. Alaa, Mahmoud H. Ismail and Hazim Tawfik}
\IEEEauthorblockA{Department of Electronics and Communications Engineering,\\
Faculty of Engineering, Cairo University,
Giza 12613, Egypt.\\ Emails: ahmedalaa@ieee.org and \{mismail, htawfik\}@eece.cu.edu.eg}}
\maketitle
\begin{abstract}
We investigate a spectrum-sharing system with non-severely faded mutual interference links, where both the secondary-to-primary and primary-to-secondary channels have a Line-of-Sight (LoS) component. Based on a Rician model for the LoS channels, we show, analytically and numerically, that LoS interference hinders the achievable secondary user capacity. This is caused by the poor dynamic range of the interference channels fluctuations when a dominant LoS component exists. In order to improve the capacity of such system, we propose the usage of an Electronically Steerable Parasitic Array Radiator (ESPAR) antenna at the secondary terminals. An ESPAR antenna requires a single RF chain and has a reconfigurable radiation pattern that is controlled by assigning arbitrary weights to $M$ orthonormal basis radiation patterns. By viewing these orthonormal patterns as multiple {\it virtual dumb antennas}, we randomly vary their weights over time creating artificial channel fluctuations that can perfectly eliminate the undesired impact of LoS interference. Because the proposed scheme uses a single RF chain, it is well suited for compact and low cost mobile terminals.           	
\end{abstract}
\IEEEpeerreviewmaketitle
\section{Introduction}
Significant interest has recently been devoted to the capacity analysis of spectrum-sharing systems in fading environments. In a spectrum-sharing system, a Secondary User (SU) transmits its data over the Primary User (PU) channel while keeping the interference experienced by the PU below a predefined level \cite{1}. Previous works have shown that fading can be exploited to improve the SU capacity when the Channel State Information (CSI) is available at the SU transmitter \cite{2}. This capacity improvement is attributed to the ability of the SU to transmit with very high power when the interference channel is severely faded. In \cite{3}, Musavian {\it et al.} derived the ergodic, outage, and minimum-rate capacities under peak and average interference constraints at the PU receiver. However, the interference experienced by the SU receiver due to primary transmission was not considered. Recently, the ergodic capacity of spectrum-sharing taking both PU and SU interference into consideration was calculated in \cite{5}. However, the analysis therein is limited to the case when all channels are severely faded, and considers an average interference power constraint only.

In a practical spectrum-sharing system, the SU capacity generally depends on two interference channels, namely; the primary-to-secondary and secondary-to-primary channels \cite{5}. If these channels are subject to Non-Line-of-Sight (NLoS) fading, then the interference channel gains perceived by the PU and SU receivers fluctuate drastically over time \cite{6}. Therefore, the SU transmitter can exploit such fluctuations by opportunistically allocating higher power to time instants when the Signal-to-Interference-and-Noise-Ratio (SINR) is large, and lower power to time instants with poor SINR \cite{2}-\cite{5}. In this paper, we study a spectrum-sharing environment with joint peak and average interference power constraints where the mutual interference channels have dominant LoS components. We show that in this case, there are limited opportunities for the SU to exploit due to the poor dynamic range of channels fluctuations, thus causing a significant capacity degradation.

Traditional multiple antenna diversity techniques were employed in \cite{9} to improve the SU capacity in spectrum sharing systems. However, the usage of multiple antennas is inhibited by the space limitations of mobile SU transceivers. This is in addition to the need for multiple RF chains, which increases the cost and complexity of the SU equipment. While such overhead is bearable for a base station, it can not be tolerated for modern mobile terminals. Moreover, emerging spectrum-sharing technologies, such as underlay {\it Device-to-Device Communications} in cellular systems, involves a mobile SU transmitter and a mobile SU receiver \cite{11}, which prevents the deployment of multiple antennas at either terminals.

The main contribution of this paper is the usage of single Electronically Steerable Parasitic Array Radiator (ESPAR) antennas at the SU transmitter and receiver in order to create artificial interference channels fluctuations, to restore the transmission opportunities limited by the dominant LoS components. This is achieved by a technique that we refer to as {\it Random Ariel Precoding} (RAP), where random time-varying complex weights are assigned to the orthonormal basis radiation patterns of the ESPAR antenna. Inspired by the seminal work of Viswanath \textit{et al} \cite{6}, we term the orthonormal radiation patterns provided by the ESPAR antenna as {\it dumb basis patterns}, since they represent Degrees of Freedom (DoFs) that are neither used to achieve diversity nor to realize multiplexing. Analytical and numerical results show that the proposed scheme can make the LoS interference transparent to both the secondary and primary systems, thus achieving the same capacity of the symmetric Rayleigh channel. The proposed scheme has several advantages. First, it does not entail extra hardware complexity as it uses a single RF chain. Second, it requires only overall CSI. Finally, it can be incorporated into low cost mobile terminals with tight space limitations.


The rest of the paper is organized as follows: the system model is presented in Section II. In Section III, the capacity degradation in the LoS interference scenario is quantified. Spectrum sharing based on the ESPAR antenna is then proposed in Section IV. Numerical results are presented in Section V and conclusions are drawn in Section VI.

\section{System Model}

\subsection{The ESPAR Antenna}
As shown in Fig. 1, an ESPAR with $M$ elements is composed of a single active element (e.g., a $\frac{\lambda}{2}$ dipole) that is surrounded by $M-1$ identical parasitic elements. Unlike multi-antenna systems, the parasitic elements are placed relatively close to the active elements. Hence, mutual coupling between different elements takes place and current is induced in all parasitic elements. The radiation pattern of the ESPAR is altered by tuning a set of $M-1$ reactive loads (varactors) $\mathbf{x} = \left[jX_{1} \ldots jX_{M-1}\right]$ attached to the parasitic elements \cite{13}. The currents in the parasitic and active elements are represented by an $M \times 1$ vector $\mathbf{i} = v_{s} (\mathbf{Y}^{-1}+\mathbf{X})^{-1}\mathbf{u}$, where $\mathbf{Y}$ is the $M \times M$ admittance matrix with $y_{ij}$ being the mutual admittance between the $i^{th}$ and $j^{th}$ elements. The load matrix $\mathbf{X}$ = {\bf diag}$\left(50 \,\,\, \mathbf{x}\right)$\footnote{The opertaion {\bf Y = diag(x)} embeds a vector {\bf x} in the diagonal matrix {\bf Y}.} controls the ESPAR beamforming, $\mathbf{u} = \left[1 \,\, 0 \ldots 0\right]^{T}$ is an $M \times 1$ vector and $v_{s}$ is the complex feeding at the active element \cite{13}. The radiation pattern of the ESPAR at an angle $\theta$ is thus given by $P(\theta) = \mathbf{i}^{T}\mathbf{a}(\theta)$, where $\mathbf{a}(\theta) = \left[a_{0}(\theta) \ldots a_{M-1}(\theta)\right]^{T}$ is the steering vector of the ESPAR at an angle $\theta$ \cite{13}-\cite{16}. The beamspace domain is a signal space where any radiation pattern can be represented as a point in this space. To represent the radiation pattern $P(\theta)$ in the beamspace domain, the steering vector $\mathbf{a}(\theta)$ is decomposed into a linear combination of a set of orthonormal basis patterns $\{\Phi_{i}(\theta)\}_{i=0}^{N-1}$ using Gram-Schmidt orthonormalization, where $N \leq M$ \cite{16}. It can be shown that the orthonormal basis patterns of the ESPAR (also known as the Ariel DoF \cite{13}-\cite{23}) are equal to the number of parasitic elements (i.e., $N = M$). Therefore, the ESPAR radiation pattern in terms of the orthonormal basis patterns can be written as \cite{13}
\begin{equation}
\label{1}
P(\theta) = \sum_{n=0}^{M-1} w_{n} \Phi_{n}(\theta),
\end{equation}
where $w_{n} = \mathbf{i}^{T}\mathbf{q}_{n}$ are the weights assigned to the basis patterns and $\mathbf{q}_{n}$ is an $M \times 1$ vector of projections of all the steering vectors on $\Phi_{n}(\theta)$. Thus, the ESPAR radiation pattern is formed by manipulating the reactive loads attached to the parasitic elements.

\subsection{Spectrum-Sharing Signal Model}
Assume a spectrum sharing model where single-user primary and secondary systems coexist as shown in Fig. 2. The received signals at the $k^{th}$ time instant are given by
\begin{equation}
\label{2}
\begin{array}{lcl} \mbox{At PU-Rx:} \,\,\, r_{p}(k) = h_{p}(k) x_{p}(k) + h_{sp}(k) x_{s}(k) + n_{p}(k),\\ \mbox{At SU-Rx:} \,\,\, r_{s}(k) = h_{s}(k) x_{s}(k) + h_{ps}(k) x_{p}(k) + n_{s}(k), \end{array}
\end{equation}
 where $h_{p}(k)$ (PU-to-PU), $h_{ps}(k)$ (PU-to-SU), $h_{sp}(k)$ (SU-to-PU) and $h_{s}(k)$ (SU-to-SU) are the respective complex-valued overall channel gains (linear combination of channel gains at all basis patterns). The primary and secondary signals $x_{p}(k)$ and $x_{s}(k)$ are complex-valued symbols drawn from an $L$-ary constellation, while $n_{s}(k)$ and $n_{p}(k)$ are AWGN samples with power spectral density $N_{o}$. The PU terminals are assumed to be equipped with single omnidirectional antennas while the antennas provided at the SU terminals are assumed to have $M$ orthonormal basis patterns having a complex weight vector $\mathbf{w} = \mathbf{i}^{T}\mathbf{q}$, where the $n^{th}$ weight $w_{n}$ has an arbitrary complex value $\sqrt{\alpha_{n}} e^{j \theta_{n}}$. The weight value depends on the setting of the parasitic elements reactive loads. This model can be reduced to the conventional single antenna (with no parasitic elements) by setting the weight vector to $\mathbf{w} = \left[1 \,\,\, 0 \ldots 0\right]$, in which case one Ariel DoF is available, corresponding to the active antenna element. We also assume perfect knowledge of these channels at the SU transmitter and receiver. It is worth mentioning that our analysis fits any design for reconfigurable antennas with beamforming capabilities, and not only the ESPAR antenna\footnote{A comprehensive framework for single-radio reconfigurable antenna design and analysis can be found in \cite{23}.}.

\section{LoS mutual interference: A hindrance to spectrum sharing capacity}
We start with the conventional single antenna scheme with all parties employing single omnidirectional antennas. In this case, we set the weight vector for the model in Section II to $\mathbf{w} = \left[1 \,\,\, 0 \ldots 0\right]$.

\subsection{Ergodic Capacity Formulation}
The ergodic capacity maximization problem is a generalization for the problems in \cite{2}, \cite{3} and \cite{5}. We adopt a joint peak and average interference power constraints, in addition to considering all the interference channels (only SU-to-PU interference is considered in \cite{2} and \cite{3}). The problem can be formulated as
\begin{align}
\label{3}
 C &= \max_{\mathbf{P_{s}}(\mathbf{\Gamma})} \mathbb{E}_{\mathbf{\Gamma}}\left\{\log_{2}\left(1+\frac{\gamma_{s}\mathbf{P_{s}}(\mathbf{\Gamma})}{\gamma_{ps}\overline{\gamma}_{p} + N_{o}}\right)\right\},\nonumber\\
&\mbox{subject to} \,\,\,\, \mathbb{E}_{\mathbf{\Gamma}} \left\{\gamma_{sp}\mathbf{P_{s}}(\mathbf{\Gamma})\right\} \leq Q_{av},\\
& \,\,\,\,\,\mbox{and} \,\,\,\, {\bf P_{s}}(\mathbf{\Gamma}) \gamma_{sp} \leq Q_{p},\nonumber
\end{align}
where $\mathbf{\Gamma} = (\gamma_{s},\gamma_{sp},\gamma_{ps}) = (|h_{s}|^{2},|h_{sp}|^{2},|h_{ps}|^{2})$, $Q_{av}$ and $Q_{p}$ are the average and peak interference power constraints, respectively, $\mathbb{E}\{.\}$ is the expectation operator, and ${\bf P_{s}}(\mathbf{\Gamma})$ is the SU transmit power allocation as a function of the CSI vector $\mathbf{\Gamma}$. Without loss of generality, we set the noise spectral density $N_{o}$ = 1 W/Hz. Thus the SNR of any link is equal to the transmit power of the PU/SU transmitter multiplied by the channel power. The PU is assumed to transmit with a constant power of $\overline{\gamma}_{p}$. The optimization problem in (\ref{3}) can be easily solved using the Lagrangian method \cite{3}. Eq. (8) in \cite{3} provides the optimal SU power allocation but without considering the PU-to-SU interference. The solution of (\ref{3}) can be obtained directly by adding the PU-to-SU interference term $\gamma_{ps}\overline{\gamma}_{p}$ to the noise variance in [4, Eq. (8)] as follows
\begin{equation}
\label{4}
{\bf P_{s}}(\mathbf{\Gamma}) = \left\{
  \begin{array}{lr}
   \frac{Q_{p}}{\gamma_{sp}}, \,\,\,\,\,\,\,\,\,\,\,\, \frac{\gamma_{sp}}{\gamma_{s}} \leq \frac{(1+\gamma_{ps}\overline{\gamma}_{p})}{\frac{1}{\lambda \log(2)}-Q_{p}}\\
    \frac{1}{\lambda \gamma_{sp} \log(2)}-\frac{\gamma_{ps}\overline{\gamma}_{p}+1}{\gamma_{s}},\\ \,\,\,\,\,\,\,\,\,\,\,\,\,\,\,\,\,\,\,\,\,\,\,\, \frac{(1+\gamma_{ps}\overline{\gamma}_{p})}{\frac{1}{\lambda \log(2)}-Q_{p}} \leq \frac{\gamma_{sp}}{\gamma_{s}} \leq \lambda \log(2) (1+\gamma_{ps}\overline{\gamma}_{p}) \\
		0, \,\,\,\,\,\,\,\, \frac{\gamma_{sp}}{\gamma_{s}} \geq \lambda \log(2) (1+\gamma_{ps}\overline{\gamma}_{p})
  \end{array}
\right.
\end{equation}
where $\lambda$ is the Lagrange multiplier and is obtained numerically to satisfy the constraints in (\ref{3}). Note that the peak interference constraint $Q_{p}$ must be greater than the average interference constraint $Q_{av}$. Thus, the peak constraint has an impact on SU capacity only if $Q_{p} > \frac{1}{\lambda \log(2)}$. If the average constraint is very large (i.e., $\lambda$ is very small), then defining a peak interference constraint is meaningless. Based on (\ref{4}), the ergodic capacity is given in (\ref{5}), where $z = \frac{\gamma_{s}}{\gamma_{sp}}$, $f_{z}(z)$ is the probability density function (pdf) of $z$, and $\{x\}^{+} = \max\{x,0\}$. In the following subsections, we investigate several channel combinations for SU-to-SU and PU-to-SU/SU-to-PU  links, highlighting the negative impact of LoS interference on SU capacity. We denote the $X-Y$ scenario as the scenario where the SU-to-SU channel is subject to a fading distribution $Y$ and the mutual interference channels are subject to fading distribution $X$. 

\begin{figure*}[!t]
\normalsize
\setcounter{mytempeqncnt}{\value{equation}}
\setcounter{equation}{4}
\begin{equation}
\label{5}
C = \mathbb{E}_{\gamma_{ps}} \left\{ \int_{z = \lambda \log(2) (1+\gamma_{ps}\overline{\gamma}_{p})}^{\frac{(1+\gamma_{ps}\overline{\gamma}_{p})}{\left\{\frac{1}{\lambda \log(2)}-Q_{p}\right\}^{+}}} \log_{2}\left(\frac{z}{\lambda \log(2)(\gamma_{ps}\overline{\gamma}_{p}+1)}\right) f_{z}(z) dz + \int_{z = \frac{(1+\gamma_{ps}\overline{\gamma}_{p})}{\left\{\frac{1}{\lambda \log(2)}-Q_{p}\right\}^{+}}}^{\infty} \log_{2}\left(1+\frac{Q_{p}\,z}{(\gamma_{ps}\overline{\gamma}_{p}+1)}\right) f_{z}(z) dz \right\}.
\end{equation}
\setcounter{equation}{\value{mytempeqncnt}+1}
\hrulefill
\vspace*{4pt}
\end{figure*}

\subsection{The Rician-Rician Scenario}
In this scenario, all the channels have LoS components. In other words, both the SU-to-SU and the mutual interference channels are not severely faded. We adopt a Rician model for the secondary and interference channels as follows:
\begin{equation}
\label{6}
h_{i}(k) = \sqrt{\overline{\gamma}_{i}}\left(\sqrt{\frac{K_{i}}{K_{i}+1}} e^{j \phi_{i}} + v_{i}(k)\right)
\end{equation}
where $h_{i}(k) \in \{h_{s}(k),h_{sp}(k),h_{ps}(k)\}$, $K_{i} \in \{K_{s},K_{sp},K_{ps}\}$, $v_{i}(k) \in \{v_{s}(k),v_{sp}(k),v_{ps}(k)\}$, and $\phi_{i} \in \{\phi_{s},\phi_{sp},\phi_{ps}\}$. The parameters $K_{s}$, $K_{sp}$, and $K_{ps}$ are the $K$-factors (specular components) of the $h_{s}$, $h_{sp}$, and $h_{ps}$ channels, respectively. The components $v_{s}(k)$, $v_{sp}(k)$, and $v_{ps}(k)$ are the diffused components where $\left(v_{s}(k),v_{sp}(k),v_{ps}(k)\right) \sim \mathcal{CN}\left(0,\left(\frac{1}{K_{s}+1}, \frac{1}{K_{sp}+1}, \frac{1}{K_{ps}+1}\right)\right)$, while ${\phi_{s}, \phi_{sp}, \phi_{ps}}$ are the constant phases of the LoS components. Non-identical Rician channels with average channel powers of $(\overline{\gamma}_{s},\overline{\gamma}_{ps},\overline{\gamma}_{sp})$ are assumed for the channels $(h_{s},h_{ps},h_{sp})$. The pdf of the channel power $\gamma_{i} = |h_{i}|^{2}$ when $|h_{i}|$ follows a Rician distribution is given by \cite{19}
\begin{equation}
\label{7}
f_{\gamma_{i}}(\gamma_{i}) = \frac{1+K_{i}}{\overline{\gamma}_{i}} e^{-K_{i}-\frac{(1+K_{i})}{\overline{\gamma}_{i}} \gamma_{i}} I_{o}\left(2 \sqrt{\frac{K_{i}(1+K_{i})}{\overline{\gamma}_{i}}\gamma_{i}}\right)
\end{equation}
where $I_{o}(.)$ is the modified Bessel function of the first kind. In the Rician-Rician scenario, both $\gamma_{sp}$ and $\gamma_{ps}$ follow the distribution in (\ref{7}). The dynamic range of the interference channels ($|h_{sp}|$,$|h_{ps}|$) fluctuations can be expressed by the variance ($\sigma^{2}_{sp,ps}$) of the Rician distribution $\sigma^{2}_{sp,ps} = 2\frac{\overline{\gamma}_{sp,ps}}{K_{sp,ps}+1} + \frac{K_{sp,ps}}{K_{sp,ps}+1} - \frac{\pi \overline{\gamma}_{sp,ps}}{2(K_{sp,ps}+1)} L_{1/2}^{2}\left(\frac{-K_{sp,ps}}{2\overline{\gamma}_{sp,ps}}\right)$ \cite{19}, where $L_{1/2}^{2}(.)$ is the square of the Laguerre polynomial. An interesting scenario is when the $K$-factor tends to $\infty$. Given that the limit of the Laguerre polynomial is $\lim_{x \to -\infty} L_{v}(x) = \frac{|x|^{v}}{\Gamma(v+1)}$, where $\Gamma(.)$ is the gamma function \cite{19}, it can be easily shown that the variance of the interference channels $\sigma^{2}_{sp,ps}$ tends to 0 when $K_{sp,ps} \to \infty$. Thus, a dominant LoS interference signal makes the interference channel almost deterministic. Therefore, for the Rician-Rician scenario with $K_{s,sp,ps} \to \infty$, the pdfs of $z$ and $\gamma_{ps}$ tend to $f_{z}(z) \to \delta(z-\frac{\overline{\gamma}_{s}}{\overline{\gamma}_{sp}})$ and $f_{\gamma_{ps}}(\gamma_{ps}) \to \delta(\gamma_{ps}-\overline{\gamma}_{ps})$. The ergodic capacity is obtained by plugging $f_{z}(z)$ and $f_{\gamma_{ps}}(\gamma_{ps})$ in (\ref{5}), leading to the same capacity of the AWGN interference channel under receive power constraint. Therefore, the capacity gain resulting from severely faded interference channels is hindered by LoS interference.

\subsection{The Rician-Rayleigh Scenario}
In this scenario, the SU-to-SU channel is characterized by a large dynamic range (large variance) and follows a Rayleigh distribution, while the interference channels have a LoS component and follow a Rician distribution. The random variable $z = \frac{\gamma_{s}}{\gamma_{sp}}$ is the ratio between the squares of a Rayleigh and a Rician random variable. In order to obtain the capacity, we first rewrite the pdf of $\gamma_{sp}$ in (\ref{7}) in terms of the Meijer-$G$ function $\MeijerG{m}{n}{p}{q}{a_1,\ldots,a_p}{b_1,\ldots,b_q}{z}$ [13, Sec. 7.8] as
\[f_{\gamma_{sp}}(\gamma_{sp}) =\]
\begin{equation}
\label{8}
 \frac{1+K_{sp}}{\overline{\gamma}_{sp}} e^{-K_{sp}-\frac{(1+K_{sp})\gamma_{sp}}{\overline{\gamma}_{sp}}} \MeijerG{1}{0}{0}{2}{-}{0, 0}{\frac{K_{sp} (1+K_{sp}) \gamma_{sp}}{\overline{\gamma}_{sp}}}.
\end{equation}
Noting that $\gamma_{s}$ follows an exponential distribution with $f_{\gamma_{s}}(\gamma_{s}) = \frac{1}{\overline{\gamma}_{s}} e^{-\frac{\gamma_{s}}{\overline{\gamma}_{s}}}$ and that the pdf of $Z = \frac{X}{Y}$ is given by $f_{Z}(z) = \int_{-\infty}^{+\infty}|y| p_{x,y}(zy, y) dy$ \cite{19}, one can obtain the expression in (\ref{9}) for the pdf of $z$. Using the property $z \MeijerG{m}{n}{p}{q}{a_1,\ldots,a_p}{b_1,\ldots,b_q}{z} = \MeijerG{m}{n}{p}{q}{a_1+1,\ldots,a_p+1}{b_1+1,\ldots,b_q+1}{z}$, the integral can be evaluated as the standard laplace transform of a Meijer-$G$ function \cite{20} yielding the result in (10). We notice that when the $K$-factors $\to\infty$, the pdfs of interest tend to
\[\lim_{K_{sp} \to \infty} f_{z}(z) = \frac{\overline{\gamma}_{sp}}{\overline{\gamma}_{s}} e^{\frac{-\overline{\gamma}_{sp}z}{\overline{\gamma}_{s}}} ,  \,\, \lim_{K_{ps} \to \infty} f_{\gamma_{ps}}(\gamma_{ps}) = \delta(\gamma_{ps} - \overline{\gamma}_{ps}).\]
Thus, the ergodic capacity of the SU is obtained by plugging the pdfs $f_{z}(z)$ and $f_{\gamma_{ps}}(\gamma_{ps})$ in (\ref{5}). Unlike the Rician-Rician scenario, the variable $z$ is not deterministic at large values of $K_{sp}$. Instead, $z$ is exponentially distributed, which means that $f_{z}(z)$ has an exponentially-bounded tail. Because a pdf with an exponentially-decaying tail gives small weight to large values of $z$, the capacity of the SU is still limited as it depends on the integral $\int_{z=0}^{\infty}\log(z)f_{z}(z) dz$. This can be physically interpreted by the fact that LoS interference channel fluctuations has a small dynamic range and thus offers limited opportunities for SU power allocation.
\begin{figure*}[!t]
\normalsize
\setcounter{mytempeqncnt}{\value{equation}}
\setcounter{equation}{8}
\begin{align}
\label{9}
f_{z}(z) &= \frac{1+K_{sp}}{\overline{\gamma}_{sp} \overline{\gamma}_{s}} \,\,\, e^{-K_{sp}} \int_{0}^{\infty} \gamma_{sp} e^{-\left(\frac{(1+K_{sp})}{\overline{\gamma}_{sp}}+\frac{z}{ \overline{\gamma}_{s}}\right) \gamma_{sp}} \MeijerG{1}{0}{0}{2}{-}{0, 0}{\frac{K_{sp} (1+K_{sp})\gamma_{sp}}{\overline{\gamma}_{sp}}} d \gamma_{sp} \\
&=\frac{\overline{\gamma}_{sp}}{\overline{\gamma}_{s}} \frac{(1+K_{sp})^{3}+z (1+K_{sp})  \frac{\overline{\gamma}_{sp}}{\overline{\gamma}_{s}}}{\left((1+K_{sp})+z \frac{\overline{\gamma}_{sp}}{\overline{\gamma}_{s}}\right)^{3}} \exp\left({\frac{-K_{sp} z \frac{\overline{\gamma}_{sp}}{\overline{\gamma}_{s}}}{(1+K_{sp})+z \frac{\overline{\gamma}_{sp}}{\overline{\gamma}_{s}}}}\right).
\end{align}
\setcounter{equation}{\value{mytempeqncnt}+2}
\hrulefill
\vspace*{4pt}
\end{figure*}
\subsection{The Rayleigh-Rayleigh Scenario}
In this scenario, all channels are severely faded with drastic fluctuations over time. Because $h_{sp}$ does not have a LoS component, the pdf of $z$ can be obtained by setting $K_{sp}$ to 0 in (10) yielding $f_{z}(z) = \frac{\overline{\gamma}_{sp}}{\overline{\gamma}_{s}\left(1+z \frac{\overline{\gamma}_{sp}}{\overline{\gamma}_{s}}\right)^{2}},$ which is the log-logistic distribution. It can be shown that the log-logistic distribution is a {\it fat-tailed distribution} by showing that $P[Z>z] \sim z^{-\alpha}$ as $z \to \infty$, where $\alpha > 0$ and $\sim$ is the asymptotic equivalence. Thus, the pdf of the variable $z$ has a heavier right-tail in the Rayleigh-Rayleigh scenario compared to the Rician-Rician and Rician-Rayleigh scenarios, which means that $\int_{z=0}^{\infty}\log(z)f_{z}(z)$ in (\ref{5}) will be larger in the Rayleigh-Rayleigh scenario as $f_{z}(z)$ is slowly decaying and will give significantly larger weights to larger values of $\log(z)$ (note that $\log(z)$ is a monotonically increasing function of $z$). On the other hand, $\gamma_{ps}$ still has an exponentially bounded tail. Again, this contributes to the SU capacity enhancement as $C$ depends on $-\int_{\gamma_{ps}=0}^{\infty}\log(\gamma_{ps}+v)f_{\gamma_{ps}}(\gamma_{ps})$ in (\ref{5}), where $v$ is a constant. Thus, severely faded interference channels offer SU capacity gain even if the SU-to-SU channel is severely faded as well. The SU opportunistic behavior can be quantified by the pdf tails for $z$ and $\gamma_{ps}$. The heavier the tail of $f_{z}(z)$ and the faster the pdf of $\gamma_{ps}$ decays, the larger is the SU capacity.

\section{Opportunistic Spectrum Sharing using Dumb Basis Patterns}
Motivated by the analysis in the previous section, we propose a technique that can eliminate the impact of LoS interference by improving the dynamic range of interference channels fluctuations. This is achieved by {\it Random Ariel Precoding} (RAP), which intentionally induces artificial fluctuations in these channels by randomizing the complex weights assigned to the basis patterns of an ESPAR antenna. Throughout this section, we adopt the system model presented in Section II, with an ESPAR antenna weight vector of $\mathbf{w} = \mathbf{i}^{T}\mathbf{q} = \left[\begin{array}{c} \sqrt{\alpha_{1}} e^{j \theta_{1}} \,\, \ldots \,\, \sqrt{\alpha_{M}} e^{j \theta_{M}}\end{array}\right]$. 

\subsection{Random Ariel Precoding (RAP)}
By viewing the Ariel DoF provided by orthonormal basis patterns as {\it virtual dumb antennas} or {\it dumb basis antennas}, we adjust the reactive loads of the parasitic antenna elements such that the weights assigned to the basis patterns are randomly varied over time. The implementation of RAP in the SU receiver is different from its implementation at the transmitter as explained hereunder.

\subsubsection{RAP at the SU transmitter}
The goal of applying RAP at the SU transmitter is to induce artificial fluctuations in $\gamma_{sp}$. For the SU transmitter to send a symbol $x_{s}(k)$ to the SU receiver at time instant $k$, it selects a weight vector $\mathbf{w} = \left[x_{s}(k)\sqrt{\alpha_{1}(k)} e^{j \theta_{1}(k)} \,\, \ldots \,\, x_{s}(k) \sqrt{\alpha_{M}(k)} e^{j \theta_{M}(k)}\right]$. Without loss of generality, we set $\alpha_{i}(k) = \frac{1}{M}, \forall i$ and vary the phases $\theta_{i}(k)$ every time instant $k$ randomly based on a uniform distribution $\theta_{i}(k) \sim \mbox{Unif}(0,2\pi)$ (independent phases are selected for all basis patterns). Hence, at each time instant $k$, the SU transmitter adjusts the reactive loads such that $\mathbf{w} = \left[\frac{x_{s}(k) e^{j \theta_{1}(k)}}{\sqrt{M}} \,\, \ldots \,\, \frac{x_{s}(k) e^{j \theta_{M}(k)}}{\sqrt{M}}\right]$. We consider $\mathbf{P(\theta)} = \left[P(\theta_{1}), \, P(\theta_{2}) \ldots \,\,, P(\theta_{G})\right]^{T}$ to be a set of $G$ spatial samples of the ESPAR radiation pattern and $\mathbf{\Phi(\theta)} = \left[\Phi_{m}(\theta_{1}), \, \Phi_{m}(\theta_{2}) \ldots \,\,, \Phi_{m}(\theta_{G})\right]^{T}$ to be a set of $G$ spatial samples of the $m^{th}$ basis pattern, where the $g^{th}$ element is the spatial element at a departure angle of $\theta_{g}$. Assuming that $h_{sp,g}(k)$ is the channel gain at an angle $\theta_{g}$ and time instant $k$, the received SU signal $r_{sp}(k)$ at the PU receiver will be given by \cite{22}
\begin{align}
\label{13}
r_{sp}(k) = x_{s}(k) \sum_{m=0}^{M-1} \frac{e^{j \theta_{m}(k)}}{\sqrt{M}} \underbrace{\left(\sum_{g=0}^{G-1}h_{sp,g}^{*}(k) \Phi_{m}(\theta_{g})\right)}_{h_{sp}^{m}(k)},
\end{align}
where $h_{sp}^{m}(k)$ is the channel gain from the $m^{th}$ basis pattern to the PU receiver. Assuming that the specular components from all basis patterns to the PU have the same $K$-factor of $K_{sp}$, one can obtain $h_{sp}^{m}(k) = \sqrt{\overline{\gamma}_{sp}}\left(\sqrt{\frac{K_{sp}}{K_{sp}+1}} e^{j \phi_{sp,m}} + v_{sp}^{m}(k)\right)$. The scattered component is a complex gaussian variable $v_{sp}^{m}(k) \sim \mathcal{CN}(0,\frac{1}{K_{sp}+1})$. On the other hand, the specular component $\sqrt{\frac{K_{sp}}{K_{sp}+1}} e^{j \phi_{sp,m}}$ from the $m^{th}$ basis pattern has a deterministic time-invariant phase shift of $e^{j \phi_{sp,m}}$. The equivalent channel $\bar{h}_{sp}(k)$ obtained by weighting the orthonormal basis patterns will thus be given by
\[
\bar{h}_{sp}(k) = \sqrt{\overline{\gamma}_{sp}} \left(\underbrace{\sum_{m=0}^{M-1} \frac{e^{j \theta_{m}(k)+j \phi_{sp,m}}}{\sqrt{M}} \sqrt{\frac{K_{sp}}{K_{sp}+1}}}_{\bar{l}_{sp}(k)} + \bar{v}_{sp}(k)\right),
\]
 where $\bar{v}_{sp}(k) = \sum_{m=0}^{M-1} \frac{e^{j \theta_{m}(k)}}{\sqrt{M}} v_{sp}^{m}(k)$ is the equivalent scattered components after applying the ESPAR weights, which were shown in \cite{6} to have the same statistics of $v_{sp}^{m}(k)$, i.e., $\bar{v}_{sp}(k) \sim \mathcal{CN}\left(0, \frac{1}{K_{sp}+1}\right)$. Recall that the main goal of RAP is to induce time varying fluctuations in the LoS component. Therefore, we are interested in evaluating the pdf of the equivalent channel $\bar{l}_{sp}(k)$ resulting from LoS propagation. This channel is formed by adding the deterministic specular components perceived by the $M$ basis patterns with random phase shifts that vary with time. Using Euler identity, $\bar{l}_{sp}(k) = \sqrt{\frac{K_{sp}}{M(K_{sp}+1)}} \sum_{m=0}^{M-1} \cos(\theta_{m}(k)+\phi_{sp,m})+ j \sin(\theta_{m}(k)+\phi_{sp,m})$. Let $Y^{R} = {\bf \mathfrak{Re}}\{\bar{l}_{sp}(k)\} = \sqrt{\frac{K_{sp}}{M(K_{sp}+1)}} \sum_{m=0}^{M-1} y_{m}$ where $y_{m} = \cos(\theta_{m}(k)+\phi_{sp,m})$. It can be easily shown that when $\theta_{m}(k) \sim \mbox{Unif}(0,2\pi)$, then $(\theta_{m}(k) + \phi_{sp,m})\mbox{mod} \, 2\pi \sim \mbox{Unif}(0,2\pi)$ as $\phi_{sp,m}$ is constant. Using random variable transformation, the pdf of $y_{m}$ is given by $f_{y_{m}}(y_{m}) = \frac{1}{\pi \sqrt{1-y_{m}^{2}}}, -1 \leq y_{m} \leq 1$, with $\mathbb{E}\{y_{m}\} = 0$, and $\mathbb{E}\{(y_{m}-\mathbb{E}\{y_{m}\})^{2}\} = \frac{1}{2}$. Applying the CLT, we get $\sum_{m=0}^{M-1} y_{m} \sim \mathcal{N}(0,\frac{M}{2})$, which implies that $Y^{R} \sim \mathcal{N}\left(0,\frac{K_{sp}}{2(K_{sp}+1)}\right)$. It can be shown that $Y^{I} = {\bf \mathfrak{Im}}\{\bar{l}_{sp}(k)\}$ has the same distribution of $Y^{R}$, which means that $\bar{l}_{sp}(k) \sim \mathcal{CN}\left(0,\frac{K_{sp}}{K_{sp}+1}\right)$. Since $\bar{l}_{sp}(k) \sim \mathcal{CN}\left(0,\frac{K_{sp}}{K_{sp}+1}\right)$, and $\bar{v}_{sp}(k) \sim \mathcal{CN}\left(0,\frac{1}{K_{sp}+1}\right)$, then the equivalent channel after applying RAP will be $\bar{h}_{sp}(k) = \sqrt{\overline{\gamma}_{sp}} (\bar{v}_{sp}(k) + \bar{l}_{sp}(k)) \sim \mathcal{CN}(0,\overline{\gamma}_{sp})$. One can therefore claim that using RAP at the SU transmitter can perfectly eliminate the impact of SU-to-PU LoS interference by employing the Ariel DoF as virtual scatterers, which converts the Rician interference channel into a Rayleigh fading one.

\subsubsection{RAP at the SU receiver}
In order to induce fluctuations in the PU-to-SU interference channel $h_{ps}$, RAP must be employed by the SU receiver. To obtain the equivalent interference channel, one can use the same analysis presented in the previous case to show that the Rician interference channel can be transformed into a Rayleigh one. The only difference is that the SU receiver does not transmit symbols. Therefore, it uses an ESPAR weight vector of $\mathbf{w} = \left[ \frac{e^{j \theta_{1}(k)}}{\sqrt{M}} \,\, \ldots \,\, \frac{e^{j \theta_{M}(k)}}{\sqrt{M}}\right]$, where the set of phase shifts $\theta_{m}(k)$ are independent and uniformly distributed.

\subsection{Smart basis patterns: Maintaining the Secondary Channel Reliability using Artificial Diversity}
It is important to note that the application of RAP will induce fluctuations not only in the interference channels, but in the SU-to-SU channel as well. Consequently, if the secondary channel is originally Rayleigh-faded, RAP will not alter its statistics. However, if the secondary channel is Rician (reliable LoS channel), RAP will turn it into a severely-faded one. Therefore, the techniques explained in the previous subsection are applicable for the Rician-Rayleigh scenario, but not for the Rician-Rician scenario. If the secondary channel is a reliable LoS channel, we aim at maintaining its reliability while inducing fluctuations in the interference channels. This can be achieved by a technique that we refer to as {\it artificial receive diversity}. First, let $\phi_{s}^{ij}$ be the phase of the specular component of the secondary channel from the $i^{th}$ transmit basis pattern to the $j^{th}$ receive basis pattern. These phases are assumed to be deterministic, constant and known at the SU receiver as they depend only on the antenna structure. Now, based on the analysis in the previous subsection, the equivalent secondary channel for a transmit ESPAR weight vector $\mathbf{w}_{T} = \left[x_{s}(k)\frac{e^{j \theta_{1}^{T}(k)}}{\sqrt{M}} \,\, \ldots \,\, x_{s}(k)\frac{e^{j \theta_{M}^{T}(k)}}{\sqrt{M}}\right]$ will be given by $\bar{h}_{s}(k) = \bar{l}_{s}(k)+\bar{v}_{s}(k)$, where $\bar{v}_{s}(k) \sim \mathcal{CN}\left(0,\frac{1}{K_{s}+1}\right)$ is the scattering components (not affected by RAP), and
\[
\bar{l}_{s}(k) = \sqrt{\overline{\gamma}_{s}} \sum_{u=0}^{M-1} \underbrace{\left(\sum_{m=0}^{M-1} \frac{e^{j \theta_{m}^{T}(k) + j \phi_{s}^{mu}}}{\sqrt{M}} \sqrt{\frac{K_{s}}{K_{s}+1}}\right)}_{\bar{l}_{s,u}(k)} w_{u}^{R},
\]
where $w_{u}^{R}$ is the $u^{th}$ element of the receive ESPAR weight vector. The specular component at the $u^{th}$ receive basis $\bar{l}_{s,u}(k)$ pattern is perceived as an artificial Rayleigh channel due to the application of RAP at the SU transmitter. In order to improve the reliability of the secondary channel, we apply conventional Maximum Ratio Combining (MRC) to the set of artificial Rayleigh channels and set $w_{u}^{R} = \frac{\bar{l}_{s,u}^{*}(k)}{||\mathbf{\bar{l}_{s}(k)}||}$, where $||\mathbf{\bar{l}_{s}(k)}|| = \sqrt{|\bar{l}_{s,1}(k)|^{2}+|\bar{l}_{s,2}(k)|^{2}+...+|\bar{l}_{s,M}(k)|^{2}}$. Therefore, the contributions of the specular component in all receive basis patterns can be averaged to suppress the fluctuations induced by RAP at the transmitter. Using the law of large numbers, when the number of receive basis patterns is large, the factor $\bar{l}_{s}(k)$ tends to be deterministic (and equal to the $K$-factor), which means that the Rician channel will be reconstructed at the receiver. In this sense, the receive basis patterns are viewed as {\it smart antennas} or {\it smart basis patterns} as they are used to achieve diversity. Note that the MRC weights $w_{u}^{R}$ are deterministic as they depend on the $K$-factor and the ESPAR transmit weights. Consequently, there will be no need to carry out channel estimation for each receive basis pattern. The overall channel $\bar{h}_{s}(k) = \bar{l}_{s}(k)+\bar{v}_{s}(k)$ is the sum of a Chi-distributed random variable $\bar{l}_{s}(k)$ and the complex gaussian scattered component $\bar{v}_{s}(k)$. Thus, channel estimation is done for the overall channel $\bar{h}_{s}(k)$ only, and the SU receiver multiplies the received signal by a weight of $\frac{\bar{h}_{s}^{*}(k)}{|\bar{h}_{s}(k)|}$. It is worth mentioning that, although the receive basis patterns act as smart antennas for the SU-to-SU link, they still act as dumb antennas for the PU-to-SU link, as the ESPAR weights at the SU receiver are not selected based on the PU-to-SU CSI and fluctuations are still induced in the interference channel.

\section{Numerical results}
\begin{figure*}
\centering
\begin{minipage}[b]{.45\textwidth}
\includegraphics[width=3.25 in]{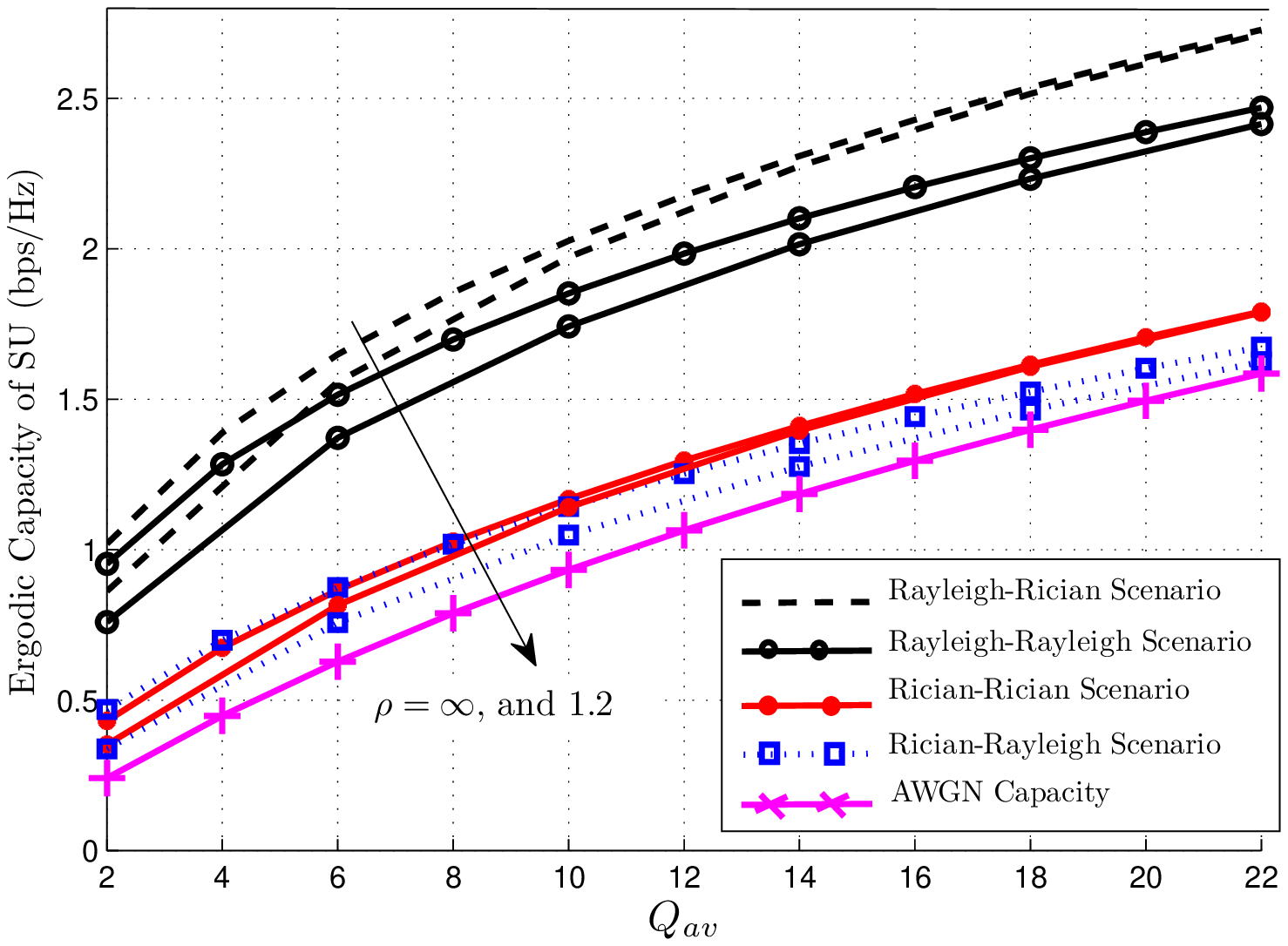}
\caption{Illustration for the impact of LoS interference on SU capacity.}
\label{ah}
\end{minipage}\qquad
\begin{minipage}[b]{.45\textwidth}
\includegraphics[width=3.25 in]{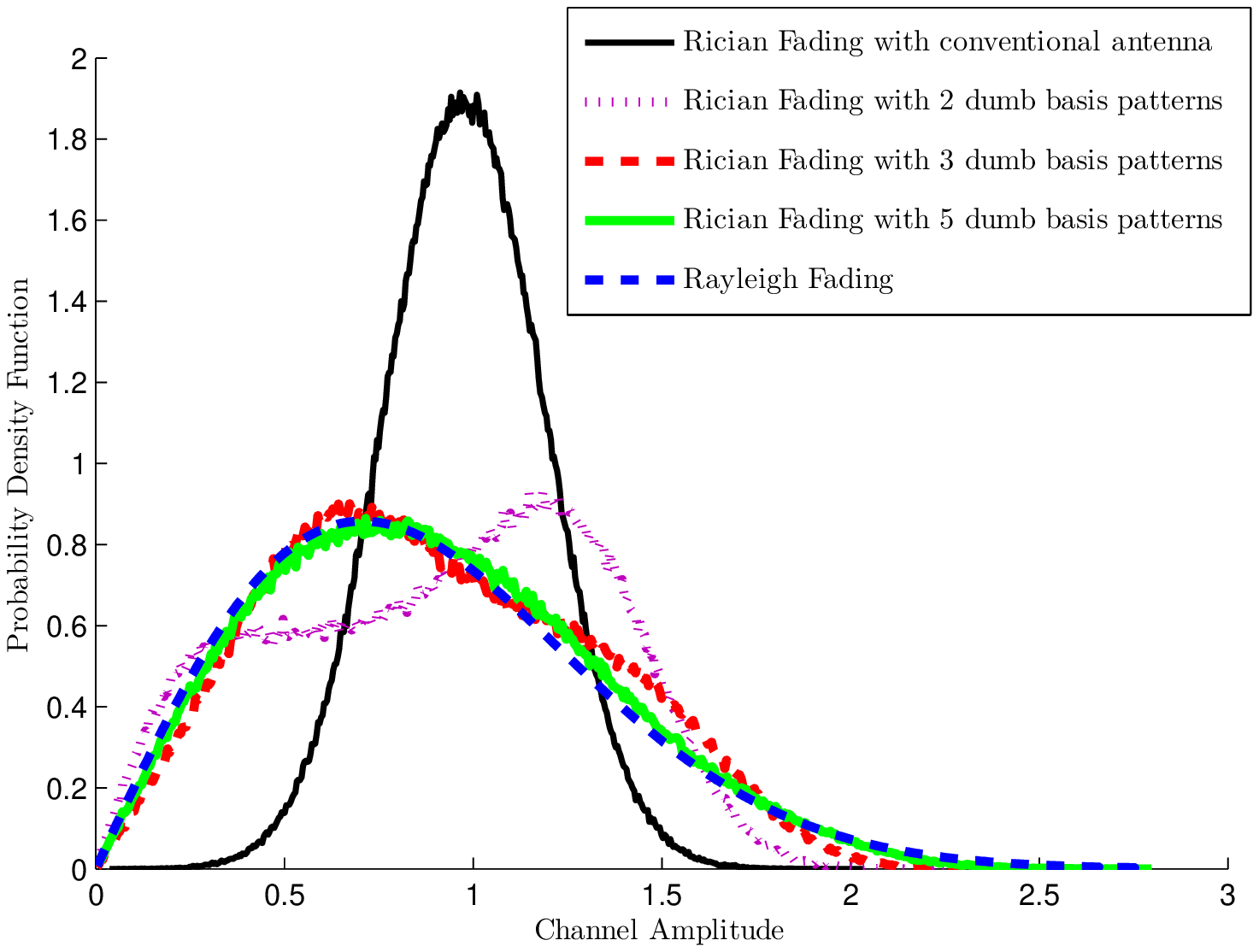}
\caption{Equivalent channel pdf after applying RAP.}
\label{ah}
\end{minipage}\qquad
\begin{minipage}[b]{.45\textwidth}
\includegraphics[width=3.25 in]{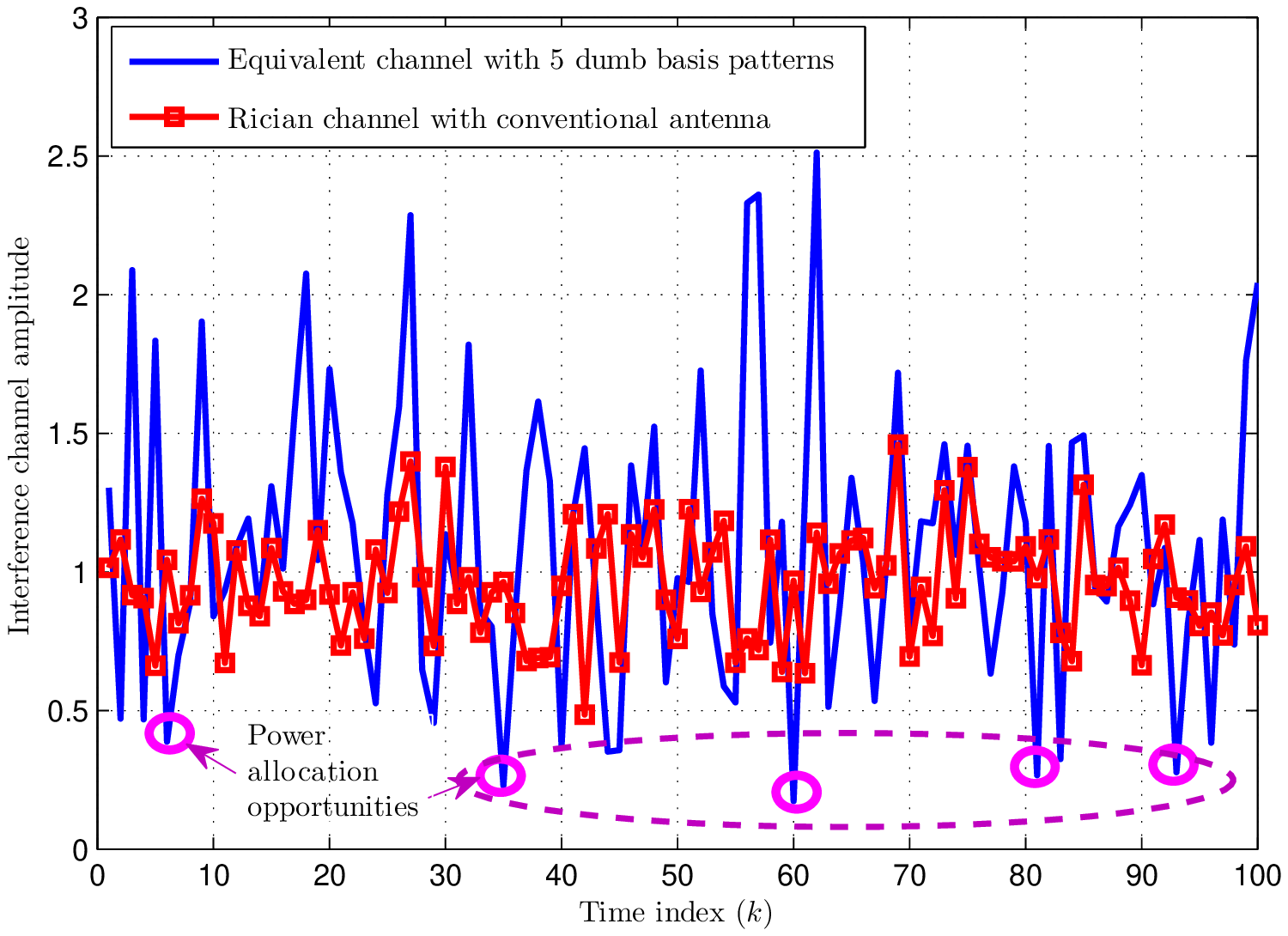}
\caption{Channel gain before and after applying RAP.}
\label{ah}
\end{minipage}\qquad
\begin{minipage}[b]{.45\textwidth}
\includegraphics[width=3.25 in]{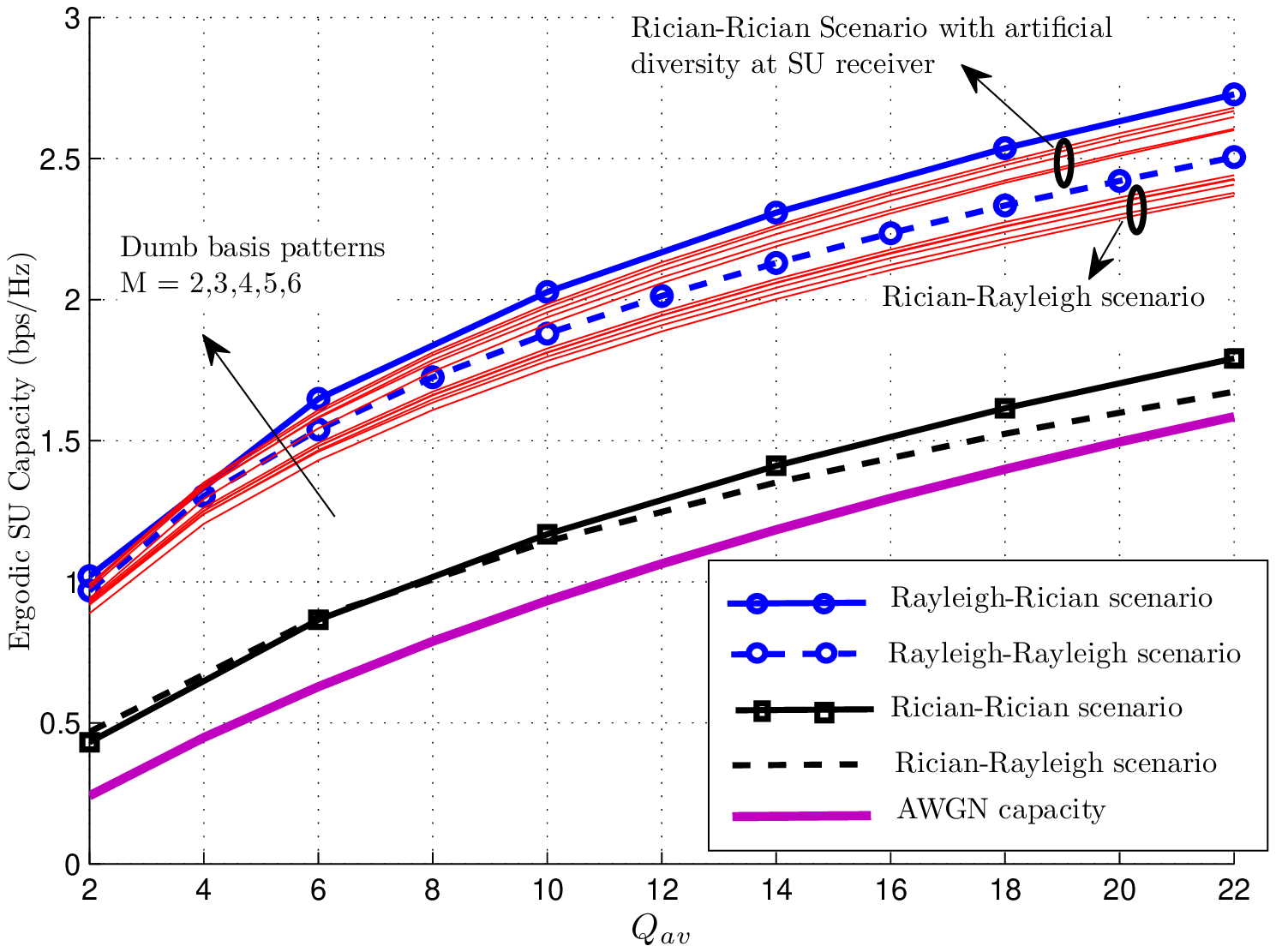}
\caption{Illustration for the impact RAP on SU capacity enhancement.}
\label{ah}
\end{minipage}
\end{figure*}

This section provides numerical results for the techniques presented throughout the paper. Monte-Carlo simulations are carried out and results are averaged over 100,000 runs. We assume the following parameter settings: $N_{o} = 1$ W/Hz, $\overline{\gamma}_{sp}=\overline{\gamma}_{ps} = \overline{\gamma}_{s} = 0$ dB, and $\overline{\gamma}_{p} = 1$ dB. For all Rician channels, we assume a $K$-factor of 10 dB. Fig. 3 depicts the ergodic capacity of the SU as a function of the average interference power constraint $Q_{av}$. We define the factor $\rho = \frac{Q_{p}}{Q_{av}}$ and plot the ergodic capacity for $\rho = \infty$ (no peak interference constraint) and $\rho = 1.2$. As expected, the SU capacity is a monotonic function of $Q_{av}$ increases, as the SU is allowed to transmit with higher power when the interference constraint is relaxed. We can also observe that when a joint peak and average interference constraint is imposed, the capacity decreases, and the amount of degradation is more significant for smaller values of $Q_{av}$.

It is notable that for all fading scenarios, the SU capacity is larger than the AWGN capacity, which agrees with the results in \cite{2}. The AWGN capacity is a special case of the Rician-Rician scenario when $K_{sp} = K_{ps} = K_{s} = \infty$. Thus, the AWGN channel is an extreme case of the LoS interference scenario tackled in Section III. Note that the Rayleigh-Rician scenario offers the best SU capacity because it enjoys a reliable SU-to-SU link, and severely-faded interference channels. LoS interference can significantly degrade the SU capacity, as a capacity gap of 1.05 bps/Hz is observed between the Rayleigh-Rician and Rician-Rician scenarios. Also, the Rayleigh-Rayleigh scenario offers a capacity increase of about 0.75 bps/Hz more than the Rician-Rician scenario, which means that the degradation caused by severely faded SU-to-SU link is less harmful than LoS interference. The Rician-Rayleigh scenario performs worse than the Rician-Rician scenario as it suffers from both severely faded SU-to-SU channel and LoS interference. It can be also observed that all fading scenarios that include a Rayleigh faded SU-to-SU channel are more sensitive to the peak interference constraint, because the large dynamic range of the SU-to-SU channel implies that the allocated SU transmit power will ``{\it hit the peak constraint}" more frequently.

To eliminate the impact of LoS interference, the concept of opportunistic spectrum sharing using dumb basis patterns was proposed. Fig. 4 shows the pdf of the interference channel amplitude ($|h_{sp}|$ or $|h_{ps}|$) before and after applying RAP for various numbers of basis patterns. It is obvious from Fig. 4 that as the number of basis patterns increases, the pdf of the equivalent channel spreads, indicating a larger dynamic range of fluctuations. It is shown that 5 basis patterns (5 parasitic elements) are enough to convert a Rician channel to a Rayleigh one. Any further increase in the number of basis patterns will not result in an increase in the dynamic range of the channel. Fig. 5 depicts the amplitude of an interference channel ($|h_{sp}|$ or $|h_{ps}|$) versus time before and after applying RAP. It is clear that after applying RAP, the resultant channel will have a larger dynamic range and more occurences of deep fades (marked with circles) than the LoS channel. In Fig. 6, we investigate the impact of the number of basis patterns on the achieved capacity gain. For the Rician-Rayleigh scenario, only one parasitic element is enough to achieve a significant capacity gain relative to the Rician-Rician scenario. Any further increase in the number of parasitic elements will make the capacity of the Rician-Rayleigh scenario converge to that of the Rayleigh-Rayleigh scenario. The same behavior is shown for the Rician-Rician scenario, where we apply the artificial diversity scheme to regain the reliability of the SU-to-SU channel.

\section{Conclusions}
In this paper, we presented a comprehensive analysis for the impact of LoS mutual interference on the SU capacity in a spectrum sharing system. It was shown that when the dynamic range of the interference channel is small, the SU capacity is significantly decreased. Stemming from this point, we introduced a novel technique to induce channel fluctuations in the interference channel using the {\it dumb basis patterns} of an ESPAR antenna. If the secondary channel contains a LoS component, we adopt an {\it artificial diversity scheme} to maintain its reliability while inducing fluctuations in the interference channels. Numerical and analytical results show that the proposed scheme can eliminate the impact of LoS interference on the SU capacity. The proposed scheme requires a single RF chain, and can fit within tight space limitations. Thus, it is adequate for low cost mobile transceivers.

\end{document}